\documentclass{appolb}
\usepackage{amsmath,fancybox}
\usepackage{epsfig,amsmath,amssymb}
\newcommand{\bea}{\begin{eqnarray}}
\newcommand{\eea}{\end{eqnarray}}
\newcommand{\ket}{{\cal i}}
\newcommand{\bra}{{\cal h}}
\newcommand{\gc}{\bra \, \frac{\alpha_s}{\pi}G^2 \, \ket}

\newcommand{\qc}{\bra \,\overline{q}q\,\ket}
\newcommand{\ga}{g_{{\cal A}}}

\begin{document}
\title{ Low Energy Aspects of  Heavy Meson Decays .%
\thanks{Presented at the Euridice meeting in Kazimierz, Poland, 24-27th of
  august 2006}%
}
\author{Jan O. Eeg
\address{Department of Physics, University of Oslo, \\
P.O.Box 1048 Blindern, N-0316 Oslo, Norway} }
\maketitle
\begin{abstract}
I discuss 
 low energy aspects  of heavy meson decays, where there is at least
 one heavy meson in the final state. Examples are
  $B -\overline{B}$ mixing,  $B  \rightarrow D \overline{D}$,
$B \rightarrow D \eta'$,  and 
 $B \rightarrow D \gamma$.
The analysis is performed in the 
  heavy quark limit within
heavy-light chiral perturbation theory. Coefficients of $1/N_c$
 suppressed chiral Lagrangian terms (beyond factorization) have been
 estimated by means of a heavy-light chiral quark model.

\end{abstract}
\PACS{PACS numbers 13.20.Hw ,  12.39.St , 12.39.Fe ,  12.39.Hg }
  
\section{Introduction}

In this paper we consider non-leptonic ``heavy meson to heavy
meson(s)'' transitions, for instance    
$B-\overline{B}$-mixing \cite{ahjoeB}, 
$B \rightarrow D \bar{D}$ \cite{EFHP} and  
  with only one $D$-meson in
 the final state, like  $B \rightarrow D \eta'$ \cite{EHP}
 and $B \rightarrow  \gamma \, D^* $ \cite{GriLe,AnVal,MacDJoe}.

The  methods \cite{BBNS} used to describe heavy to light tansitions like 
$B \rightarrow \pi \pi$ and $B \rightarrow K
\pi$ are not suited  for the decays we consider.
We use heavy-light chiral perturbation theory (HL$\chi$PT).
Lagrangian terms corresponding to factorization are then determined
to zeroth order in $1/m_Q$, where $m_Q$ is the mass of the heavy quark
($b$ or $c$).  For $B-\overline{B}$-mixing we have also calculated
$1/m_b$ corrections \cite{ahjoeB}. 

Colour suppressed  $1/N_c$ terms 
beyond factorization can be written down, but their coefficients are unknown. 
However, these coefficients can be calculated within a
 heavy-light chiral quark model 
 (HL$\chi$QM) \cite{ahjoe}
 based on the heavy quark effective theory 
(HQEFT) \cite{neu} and HL$\chi$PT \cite{itchpt}.
The $1/N_c$ suppressed non-factorizable terms calculated in this way
will typically be proportional to  a
 model dependent gluon condensate \cite{ahjoeB,EFHP,EHP,MacDJoe,ahjoe,BEF}.

\section{Quark Lagrangians for non-leptonic decays}

The effective non-leptonic Lagrangian at
quark level has the  form \cite{BuBuLa}: 
 \begin{equation}
 {\mathcal L}_{W}=  \sum_i  C_i(\mu) \; \hat{Q}_i (\mu) \; ,
\end{equation}
where  the Wilson coefficients $C_i$ contain $G_F$ and KM factors.
Typically, the operators  are four quark operators being the product
of two currents:
\begin{eqnarray}
\hat{Q}_{i} \, = \,  j_W^\mu(q_1 \to q_2) \; j^W_\mu(q_3 \to q_4) \; ,
\end{eqnarray}  
where $j_W^\mu(q_i \to q_j) = \overline{(q_j)_L} \, \gamma^\mu \,  (q_i)_L$, and 
some of the quarks $q_{i,j}$ are heavy. 
To leading order in $1/N_c$, matrix elements of $\hat{Q}_{i}$ factorize 
in products of matrix elements of currents.
Non-factorizable $1/N_c$ suppressed  terms are obtained  from
``coloured quark operators''.
Using Fierz transformations   and
\begin{equation}
\delta_{i j}\delta_{l n}  =   \frac{1}{N_c} \delta_{i n} \delta_{l j}
 \; +  \; 2 \; t_{i n}^a \; t_{l j}^a \; , 
\end{equation} 
where $t^a$ are colour matrices,
we may rewrite the operator $\hat{Q}_i$ as 
\begin{eqnarray} 
\hat{Q}_i^F = \frac{1}{N_c} \, j_W^\mu(q_1 \to q_4) \; j^W_\mu(q_3 \to q_2) 
+ 2 \,  j_W^\mu(q_1 \to q_4)^a \; j^W_\mu(q_3 \to q_2)^a \; ,
\label{ColOp}
\end{eqnarray}  
where
$j_W^\mu(q_i \to q_j)^a = \overline{(q_j)_L} \, \gamma^\mu \,  t^a \, (q_i)_L$
is a left-handed coloured current. The quark operators in $\hat{Q}_i^F$
give $1/N_c$ suppressed terms.

\section{Heavy-light chiral perturbation theory}

The QCD Lagrangian involving light and heavy quarks is:
\begin{eqnarray}
{\cal L}_{Quark}= 
\pm \overline{Q^{(\pm)}_{v}} i v \cdot D Q^{(\pm)}_{v}
 + {\cal O}(m_Q^{- 1}) \, + \, \bar{q}i\gamma \cdot  D q \; + ...
\label{LQuark}
\end{eqnarray}
where  $Q^{(\pm)}_{v}$ are the quark fields for a heavy quark and a
 heavy 
 anti-quark with velocity $v$, $q$ is the light quark triplet, and
 $i D_\mu = i \partial_\mu -e_q A_\mu - g_s t^a A_\mu^a$. The
bosonized Lagrangian  have the following form, consistent with the
underlying symmetry \cite{itchpt}:
\bea
{\cal L}_{\chi}(Bos) = 
\mp Tr\left[\overline{H_{a}^{(\pm)}}(iv\cdot {\cal D}_{fa})H_{f}^{(\pm)}\right]
- g_{\cal A}Tr\left[\overline{H_{a}^{(\pm)}}H_{f}^{(\pm)} \gamma_\mu\gamma_5 
{\cal A}^\mu_{fa}\right] + ... 
\label{StrongL}
\eea
where the covariant derivative is
 $i {\cal D}_{fa}^\mu \equiv  \delta_{af} (i \partial^\mu 
- e_H A^\mu) -{\cal V}_{fa}^\mu \; $ ; $a,f$ being SU(3) flavour
indices. 
The axial coupling is $\; \ga \simeq 0.6$.  
Furthermore,
\begin{equation}
{\cal V}_{\mu}  (\text{or} \,  {\cal A}_\mu ) \, = \, 
\pm \frac{i}{2}(\xi^\dagger\partial_\mu\xi
 \pm \xi\partial_\mu\xi^\dagger ) \; , 
\end{equation}
where   $\xi=exp(i\Pi/f)$, and $\Pi$ is a 3 by 3 matrix
containing the light mesons ($\pi, K \eta$), and
the heavy $(1^-,0^-)$ doublet field $(P_\mu,P_5)$ is
\begin{equation}
H^{(\pm)} =  P_{\pm} (P_{\mu}^{(\pm)} \gamma^\mu -     
i P_{5}^{(\pm)} \gamma_5) \; \, , \, P_{\pm} =  (1 \pm \gamma \cdot v)/2 \; ,
\end{equation}
where superscripts $(\pm)$ means meson and anti-meson respectively.
To bosonize the non-leptonic quark Lagrangian, we need to bosonize the currents. Then
the $b$, $c$, and $\overline{c}$ quarks are treated within HQEFT, which
 means the replacements
 $b \rightarrow Q_{v_b}^{(+)}  \; ,
 c \rightarrow Q_{v_c}^{(+)}  \; $,  and
 $\overline{c} \rightarrow Q_{\bar{v}}^{(-)}$.
Then the bosonization of currents within HQEFT  
for decay of a heavy  $B$-meson will be:
\begin{equation}
 \overline{q_L} \,\gamma^\mu\, Q_{v_b}^{(+)} \;  \longrightarrow \;
    \frac{\alpha_H}{2} Tr\left[\xi^{\dagger} \gamma^\mu
L \,  H_{b}^{(+)} \right] \; \equiv \; J_b^\mu \; ,
\label{Jb}
\end{equation}
where  
$L$ is the left-handed projector in Dirac space, and
 $\alpha_H =  f_H \sqrt{M_H}$ for $H=B,D$ 
before  pQCD and chiral corrections are added. 
Here,  $H_{b}^{(+)}$ represents the heavy meson (doublet) containing a $b$-quark.
 For creation of a heavy anti-meson $\overline{B}$ or $\overline{D}$,
 the corresponding currents $J_{\bar{b}}^\mu$ and  $J_{\bar{c}}^\mu$ are
 given by (\ref{Jb}) with $H_{b}^{(+)}$  replaced by 
 $H_{b}^{(-)}$ and  $H_{c}^{(-)}$, repectively.
For the  $B \rightarrow D$ transition we have
\begin{equation}
 \overline{Q_{v_b}^{(+)}} \,\gamma^\mu\, L Q_{v_c}^{(+)}\;  \longrightarrow \;
    - \zeta(\omega) Tr\left[
 \overline{H_c^{(+)}} \gamma^\mu L  H_{b}^{(+)} \right] \; \equiv \;
 J_{b \rightarrow c}^\mu \; ,
\label{Jbc}
\end{equation}
where $\zeta(\omega)$ is the Isgur-Wise function, and   $\; \omega = v_b \cdot v_c$. 
For creation of  $D \overline{D}$ pair we have the same expression for
the current  $J_{c\bar{c}}^\mu$
with  $H_{b}^{(+)}$ replaced by  $H_{c}^{(-)}$, and  $\zeta(\omega)$
replaced by $\zeta(-\lambda)$, 
where  $\lambda= \bar{v} \cdot v_c \; $. In addition there are
 $1/m_Q$ corrections for  $Q=b,c$.
 The low velocity
  limit is  $\omega \rightarrow 1 \, $. For $ \, B \rightarrow D
 \overline{D}$ and  $ \, B \rightarrow D^* \gamma$ one  has
  $\omega \simeq 1.3 \, ,$ and 
  $\omega \simeq 1.6 \, $, respectively.

\subsection{Factorized lagrangians for non-leptonic processes}

 For $B -\overline{B}$ mixing, the factorized bosonized Lagrangian is
\begin{equation}
{\mathcal L}_{B}^{} = C_B \; J_b^\mu \; (J_{\bar{b}})^\mu \; , 
\label{BBbar}
\end{equation}
where $C_B$ is a short distance Wilson coefficient (containing
 $(G_F)^2$), which is 
 taken at $\mu = \Lambda_\chi \simeq$ 1 GeV, and the currents are given by
 (\ref{Jb}).

\begin{figure}[t]
\begin{center}
   \epsfig{file=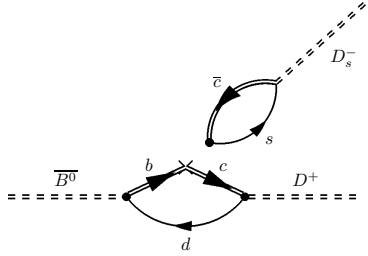,width=5.5cm}
\caption{\small{Factorized contribution for 
$\overline{B^0_d}  \rightarrow D^+ D_s^-$
through the spectator mechanism, which does not exist for
 decay mode $\overline{B^0_d} \rightarrow D_s^+ D_s^-$.}}
\label{fig:bdd_fact}
\end{center}
\end{figure}

For processes obtained from two different four quark operators for $b \rightarrow c
\bar{c} q \; \; (q=d,s)$, we find the  factorized
Lagrangian corresponding to Fig.~\ref{fig:bdd_fact}:
 \bea
 {\mathcal L}_{Fact}^{Spec}= 
 (C_2+\frac{C_1}{N_c})  J_{b \rightarrow c}^\mu \; (J_{\bar{c}})_\mu
 \; ,
 \label{FactSpec}
\eea
where $C_i = \frac{4}{\sqrt{2}} G_F  V_{cb} V_{cq}^* \, a_i$, and \cite{GKMWF}
$a_1 \simeq -0.35-0.07i$, $a_2 \simeq 1.29+0.08i$. 
We have considered the process  
$\overline{B^0_d} \rightarrow D_s^+ D_s^-  \, $. Note that
there is no factorized contribution  to this process if both
$D$-mesons in the final state are pseudoscalars! But the factorized
contribution to $\overline{B^0_d} \rightarrow D^+ D_s^- $ will
be the starting point for  chiral loop contributions to 
 the process $\overline{B^0_d} \rightarrow D_s^+ D_s^- $. 
\begin{figure}[t]
\begin{center}
   \epsfig{file=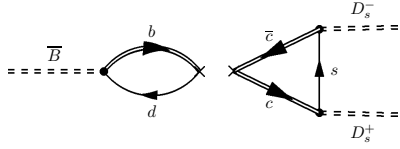,width=5.8cm}
\caption{\small{Factorized contribution for 
$\overline{B^0_d}  \rightarrow D_s^+ D_s^-$
through the annihilation  mechanism, which give zero contributions if
both $D_s^+$ and $D_s^-$ are pseudoscalars.}}
 \label{fig:bdd_fact2}
\end{center}
\end{figure}
The factorizable term from annihilation
 is shown in Fig.~\ref{fig:bdd_fact2}, and is:
\bea
{\mathcal L}_{Fact}^{Ann}= 
 (C_1+\frac{C_2}{N_c}) \,  J_{c \bar{c}}^\mu \; (J_{b})_\mu \; .
\eea
 Because $(C_1+C_2/N_c)$
is a non-favourable combination of the Wilson coefficients, this term
 will give a small non-zero contribution if at least one of 
the mesons in the final state is  a vector.

\subsection{Possible $1/N_c$ suppressed tree level terms }

For $B- \bar{B}$ mixing, we have for instance the  $1/N_c$ suppressed term
 \bea
Tr\left[\xi^{\dagger} \sigma^{\mu \alpha}
L \,  H_{b}^{(+)} \right]  \cdot 
 Tr\left[\xi^{\dagger} \sigma_{\mu \alpha} R  H_{\bar{b}}^{(-)}
 \right] \; .
\eea
For $B \rightarrow D \bar{D}$, we have for instance the terms 
 \bea
Tr\left[\xi^{\dagger} \sigma^{\mu \alpha}
L \,  H_{b}^{(+)} \right]  \cdot 
 Tr\left[
 \overline{H_c^{(+)}} \gamma_\alpha L  H_{\bar{c}}^{(-)} \gamma_\mu
 \right] \; ,
\eea
\bea
Tr\left[\xi^{\dagger} \sigma^{\mu \alpha}
L \,  H_{b}^{(+)} \right] \cdot
 Tr\left[
 \overline{H_c^{(+)}} \gamma_\alpha L  H_{\bar{c}}^{(-)} \right] 
(\bar{v}-v_c)_\mu \; \, .
\eea
One needs a framework to estimate the coefficients of such terms. We  use the 
 HL$\chi$QM, which will pick a certain linear combination of  $1/N_c$ terms.

\subsection{Chiral loops for non-leptonic processes}

Within HL$\chi$PT, the leading chiral corrections are proportional to
\bea
 \chi(M) \equiv (\frac{\ga m_M}{4 \pi f})^2
 \ln(\frac{\Lambda_\chi^2}{m_M^2}) \; ,
\eea
where $m_M$ is the appropriate light meson mass and $\Lambda_\chi$
is the chiral symmetry breaking scale, which is also the matching
scale within our framework.

\begin{figure}[t]
\begin{center}
\epsfig{file=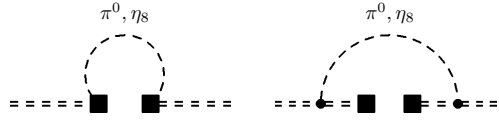,width=7cm,clip=true}
\caption{Chiral corrections to $B -\overline{B}$ mixing, i.e  the bag
  parameter $B_{B_q}$ for $q=d,s$. The black boxes are weak vertices.}
\label{fig:bagparam}
\end{center}
\end{figure}

For  $B -\overline{B}$ mixing there are chiral loops obtained from 
(\ref{StrongL}) and (\ref{BBbar}) shown
in Fig.~\ref{fig:bagparam}. These have to be added to the factorized
contribution. 
\begin{figure}[t]
\begin{center}
   \epsfig{file=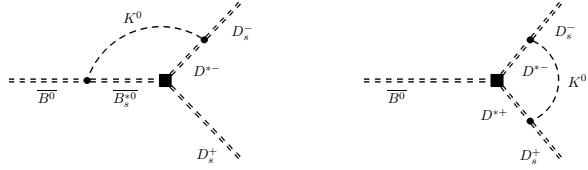,width=8cm}
\caption{Two classes of non-factorizable chiral loops for 
$\overline{B^0_d} \rightarrow D_s^+ D_s^-$ based on the factorizable amplitude
proportional to the IW function  $\sim \zeta(\omega)$.}
\label{fig:chiral1}
\end{center}
\end{figure}

For the process $\overline{B^0_d} \rightarrow D_s^+ D_s^-$ we obtain a 
chiral loop amplitude corresponding to Fig.~\ref{fig:chiral1}. This
amplitude is complex and depend on $\omega$ and $\lambda$ defined previously.
It has been recently shown \cite{AnVal} that  $(0^+,1^+)$ states in loops 
should also be added to the result.

\section{The heavy-light chiral quark model}

\begin{figure}[t]
\begin{center}
\epsfig{file=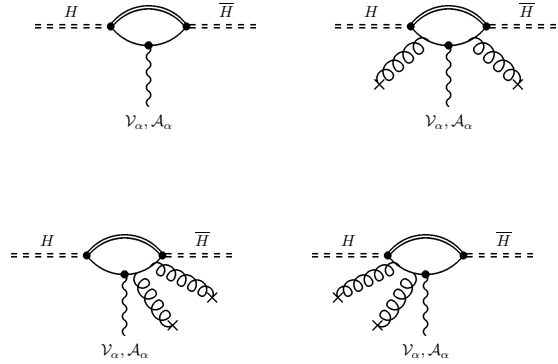,width=7cm }
\caption{Diagrams generating the strong chiral lagrangian at mesonic level.
The kinetic term and and the axial vector term $\sim \ga$.}
\label{fig:va}
\end{center}
\end{figure}

 The Lagrangian for HL$\chi$QM  \cite{ahjoe} contains the Lagrangian (\ref{LQuark}):
\begin{equation}
{\cal L}_{HL \chi QM} = {\cal L}_{HQET}+ {\cal L}_{\chi QM} + {\cal
  L}_{Int} \; , 
\label{HLChiQM}
\end{equation}
where ${\cal L}_{HQET}$ is the heavy quark part of (\ref{LQuark}), and
the light quark part is 
\begin{equation}
{\cal L}_{\chi QM} =  
\overline{\chi} \left[\gamma^\mu (i D_\mu + {\cal V}_{\mu}  +  
\gamma_5  {\cal A}_{\mu}) -  m \right]\chi \; .
\end{equation}
Here $\chi_L  =   \xi^\dagger q_L \, $ and $\, \chi_R  = \xi q_R $ are 
flavour rotated light quark fields, and $m$ is the light  constituent mass.  
\begin{figure}[t]
\begin{center}
\epsfig{file=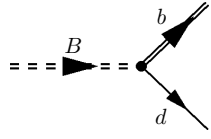,width=3.0cm }
\caption{The $HL\chi$QM ansatz: Vertex for quark meson interaction}
\label{fig:vertex}
\end{center}
\end{figure}
The bosonization of the (heavy-light) quark sector is performed via the ansatz:
\begin{equation}
{\cal L}_{Int}  =   
 -   G_H \, \left[\overline{\chi}_f \, \overline{H_{v}^{f}} \, Q_{v} \,
  +     \overline{Q_{v}} \, H_{v}^{f} \, \chi_f \right] \; . 
\label{Int}
\end{equation}
The coupling $G_H$ is determined by  bosonization through the loop
diagrams in Fig~\ref{fig:va}. The bosonization lead to relations between
the model dependent parameters $G_H$, $m$, and $\gc$, and the
quadratic-, linear, and logarithmic- divergent integrals $I_1, I_{3/2},
I_1$, and the physical  quantities $f_\pi$, $\qc$, $\ga$ and $f_H$ ($H=B,D$).
For example, the relation  obtained   for identifying the  kinetic term is:
\begin{equation}
 -  iG_H^2N_c \, (I_{3/2}  +   2mI_2  +  
\frac{i(8-3\pi)}{384 N_c m^3}\gc) =   1 \; ,
\label{Identity}
\end{equation}
where we have used the prescription:
\begin{equation}
g_s^2 G_{\mu \nu}^a G_{\alpha \beta}^a  \; \rightarrow 4 \pi^2
 \gc \frac{1}{12} (g_{\mu \alpha} g_{\nu \beta} -  
g_{\mu \beta} g_{\nu \alpha} ) \, .
\end{equation}
The parameters are fitted in strong sector, with
$ \gc= [(0.315\pm0.020)$ GeV]$^{4}  \; , $ and $  \;   {G_H}^2 = \frac{2
  m}{f^2} \rho \,$,  
 where $\rho \simeq 1$. For  details , see \cite{ahjoe}.

\section{$1/N_c$ terms from HL$\chi$QM}

\begin{figure}[t]
\begin{center}
   \epsfig{file=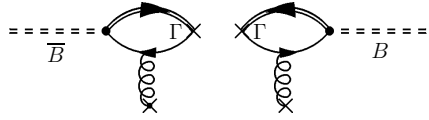,width=6cm}
\caption{Non-factorizable contribution to  $B -\overline{B}$ mixing;
 $\Gamma \equiv t^a\,\gamma^\mu\,L$}
\label{BBMixC}
\end{center}
\end{figure}
\begin{figure}[t]
\begin{center}
   \epsfig{file=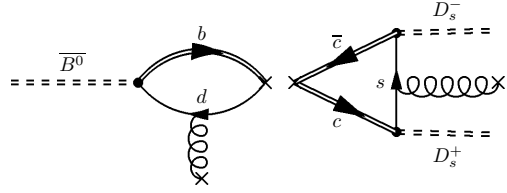,width=7cm}
\caption{Non-factorizable $1/N_c$ contribution for 
$\overline{B^0}  \rightarrow D_s^+ D_s^-$
through the annihilation mechanism with additional soft gluon emision.}
\label{fig:bdd_nfact2}
\end{center}
\end{figure}
\begin{figure}[t]
\begin{center}
\epsfig{file=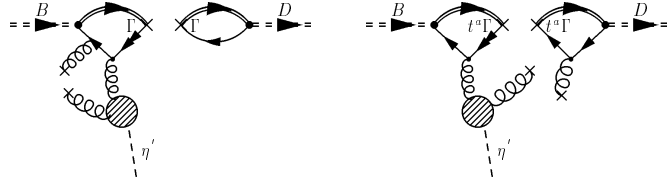,width=9cm}
\caption{Diagram for $B \rightarrow D \eta'$
within $HL\chi QM$. $\Gamma = \gamma^\mu (1-\gamma_5)$}
\label{fig:BDeta}
\end{center}
\end{figure}
\begin{figure}[t]
\begin{center}
   \epsfig{file=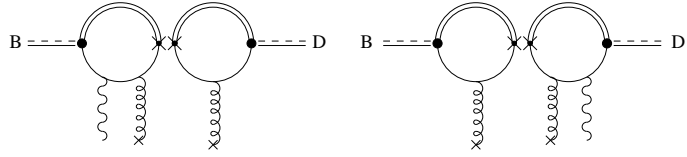,width=9cm}
\caption{Non-factorizable  contributions to $B \rightarrow \gamma D^*$
from the coloured operators}
 \label{fig:Nonfact}
\end{center}
\end{figure}

To obtain the $1/N_c$ terms for  $B -\overline{B}$ mixing in Fig.~\ref{BBMixC} ,
 we need the bosonization of colored current in the quark operators of
 eq. (\ref{ColOp}): 
\bea
\left(\overline{q_L}\, t^a  \,\gamma^\alpha \, Q_{v_b}^{(+)}\right)_{1G} 
\;   \longrightarrow
- \frac{G_H \, g_s}{64 \pi} \,G_{\mu\nu}^a
Tr\left[\xi^\dagger \gamma^\alpha  L \, H_b^{(+)} \Sigma_{\mu \nu}
\right] \; ,
\label{HLColCu}
\eea
\bea
\Sigma^{\mu\nu} = 
 \sigma^{\mu\nu} \, - 
 \frac{2 \pi f^2}{m^2 \, N_c}  [ \sigma^{\mu\nu},
 \gamma \cdot v_b ]_+ \; \, .
\eea
This coloured current is also used for 
 $B \rightarrow D \overline{D}$ in Fig.~\ref{fig:bdd_nfact2}, for 
 $B \rightarrow D \, \eta'$ in Fig.~\ref{fig:BDeta}, and for  
 $B  \rightarrow \gamma  D^{*}$ in Fig.~\ref{fig:Nonfact}
In addition there are more complicated bosonizations of coloured
currents as indicated in  Fig.~\ref{fig:bdd_nfact2}.

For $B \rightarrow D \, \eta'$ and $B  \rightarrow \gamma  D^{*}$
decays there are two different four quark operators, both
 for $b \rightarrow c \bar{u} q$ and 
 $ b \rightarrow \bar{c} u q$, respectively.
 At  $\mu=1$ GeV they have  Wilson coefficients
$a_2  \simeq 1.17 \;, \; a_1  \simeq -0.37$ (up to prefactors $G_F$
 and KM-factors).
For  $B \rightarrow D \, \eta'$, we must also
attach a propagating gluon to the $\eta' g g^*$-vertex.
Note that for $\overline{B_{s,d}^0}  \rightarrow \gamma D^{0*}$,
the $1/N_c$ suppressed mechanism in Fig.~\ref{fig:Nonfact}
dominates, unlike $\overline{B_{s,d}^0}  
\rightarrow \overline{\gamma D^{0*}}$.
Factorized  contributions are proportional to either
the favourable contribution 
 $a_f \; = \; a_{2} + a_{1}/N_c \simeq 1.05$ or the 
non-favourable contribution  
  $a_{nf} \; = \;  a_{1} + a_{2}/N_c \simeq 0.02$.

\subsection{$1/m_c$ correction  terms }

 For the $B \rightarrow D$ transition we have the $1/m_c$ suppressed terms:
\bea
\frac{1}{m_c} \, Tr\left[ \left( Z_0
 \overline{H_c^{(+)}}   + Z_1 \gamma^\alpha  \overline{H_c^{(+)}}
 \gamma_\alpha  + Z_2 \overline{H_c^{(+)}} \gamma \cdot v_b \right) 
\gamma^\alpha L  H_{b}^{(+)} \right] \; ,
\eea
where the $Z_i$'s are calculable within HL$\chi$QM.
The relative size of $1/m_c$ corrections are typically of order 
 $ 20-30 \%$.

 \section{Results}

\subsection{$B- \overline{B}$ mixing}

The result for the B(ag) parameter in $B-\overline{B}$-mixing has the
form \cite{ahjoeB}
\begin{equation}
\hat{B}_{B_q}=\frac{3}{4} \, \widetilde{b}
 \left[ 1 + \frac{1}{N_c} \left(1 - \delta_G^B
\right) + \frac{\tau_b}{m_b} + \frac{\tau_\chi}{32 \pi^2 f^2} \right]
 \; \, ,
\label{Bagparam}
\end{equation}
 similar to the  $K-\overline{K}$-mixing case \cite{BEF}.
From perturbative QCD we have $\widetilde{b} \simeq 1.56$ at $\mu= \Lambda_\chi$ =
1 GeV. From calculations within the HL$\chi$QM we obtain,
$\delta_G^B=0.5\pm0.1$ and  $\tau_b =(0.26\pm0.04)$GeV, 
and from chiral corrections $\tau_{\chi,s}=(-0.10\pm0.04)$GeV$^2$, and
 $\tau_{\chi,d}=(-0.02\pm0.01)$GeV$^2$. We obtained
\bea
\hat{B}_{B_d}=1.51\pm 0.09 \qquad \hat{B}_{B_s}=1.40 \pm 0.16 \; ,
\eea
in agreement  with lattice results. 

\subsection{$B \rightarrow D \, \overline{D}$ decays} 

Keeping the chiral logs and the $1/N_c$ terms from the gluon condensate, 
 we find the branching ratios in the  ``leading approximation''. 
For decays of $\bar{B}_d^0$ ($\sim V_{cb} V^*_{cd}$) and 
$\bar{B}_s^0$ ($\sim V_{cb} V^*_{cs}$)
we obtain branching ratios of order few $\times  10^{-4}$
and $\times  10^{-3}$, respectively
Then we have to add counterterms $\sim m_s$ for chiral loops. These may be
estimated in HL$\chi$QM.

\subsection{$B \rightarrow D \, \eta'$ and $B  \rightarrow \gamma  D^{*}$
decays} 

The result corresponding to Fig.~\ref{fig:BDeta} is:
\begin{equation}
Br(B\rightarrow D\eta^{'}) \,\simeq \, 2 \times 10^{-4} \; \, . 
\end{equation}
The partial branching ratios 
from the mechanism in Fig.~\ref{fig:Nonfact} are \cite{MacDJoe}
\begin{equation}
Br(\overline{B^0_d} \rightarrow \gamma \, D^{*0})_G \, \simeq  1 \times 10^{-5}
\quad ; \quad
Br(\overline{B^0_s} \rightarrow \gamma \, D^{*0})_G \, \simeq  6 \times 10^{-7}
\; .
\end{equation}
The corresponding factorizable  contribibutions are 
 roughly two orders of magnitude smaller.
Note that the process 
$\overline{B^0_d} \rightarrow \gamma \, \overline{D^{*0}}$ has 
substantial meson exchanges (would be chiral loops for $\omega
 \rightarrow 1$), and is different.

\section{Conclusions}

Our low energy framework is well suited to $B - \overline{B}$ mixing, and to
some extent to
 $B \rightarrow D \overline{D}$.
Work continues to include $(0^+,1^+)$, states,
 counterterms, and $1/m_c$ terms.
Note that the amplitude for 
$\overline{B^0_d} \rightarrow D_s^+ D_s^- \, $
is  zero in the factorized limit.
For processes like $B \rightarrow  D \eta'$ and 
$B \rightarrow  D \gamma \; $ we can  give order of magnitude estimates
when factorization give zero or small amplitudes.

\vspace{0.1cm}
\begin{center}
* * *
\end{center}

\vspace{0.1cm}

JOE is supported in part by the Norwegian
 research council
 and  by the European Union RTN
network, Contract No. HPRN-CT-2002-00311  (EURIDICE).
He thanks his collaborators :   A. Hiorth, S. Fajfer, A. Polosa, A.
Prapotnik Brdnik, J.A. Macdonald S\o rensen, and J. Zupan

\bibliographystyle{unsrt}

\end{document}